\documentclass[prl,twocolumn,showpacs,amssymb]{revtex4}
\usepackage{graphicx}

\begin{document}

\title{Electrostatic Modulation of the Superfluid Density in an Ultrathin La$_{2-x}$Sr$_{x}$CuO$_{4}$ Film}
\author{A. R\"{u}fenacht$^{1}$, J.-P. Locquet$^{2}$, J. Fompeyrine$^{2}$, D. Caimi$^{2}$, and P. Martinoli$^{1}$}
\affiliation{$^{1}$Institut de Physique, Universit\'{e} de Neuch\^{a}tel, CH-2000 Neuch\^{a}tel, Switzerland\\$^{2}$IBM Research Division, Zurich Research Laboratory, CH-8803 R\"{u}schlikon, Switzerland}

\date{\today}

\begin{abstract}
By capacitively charging an underdoped ultrathin
La$_{2-x}$Sr$_{x}$CuO$_{4}$ film with an electric field applied
across a gate insulator with a high dielectric constant, relative
changes of the areal superfluid density $n_{s\Box}$ of
unprecedented strength were observed in measurements of the film
kinetic inductance. Although $n_{s\Box}$ appears to be
substantially reduced by disorder, the data provide, for the first
time on the same sample, direct compelling evidence for the Uemura
relation $T_{c} \propto n_{s\Box}(T=0)$ in the underdoped regime
of copper-oxide superconductors.
\end{abstract}

\pacs{74.62.Yb, 74.78.Bz, 74.25.Nf, 74.90.+n}


\maketitle
It is widely accepted that copper-oxide superconductors in the underdoped regime are doped charge transfer insulators whose properties strongly depend on the concentration of the "free" electrical carriers. If the doping level is measured by a parameter $x$ expressing the number of free carriers per Cu site in  the CuO$_{2}$ planes, in the hole-doped materials superconductivity typically sets in at $x\approx0.05$,  reaches its maximum strength at  $x\approx0.15$, and disappears above $x\approx0.30$. Usually, $x$ is changed by non-isovalent chemical substitution of the antiferromagnetic insulating parent compound ($x=0$): for instance, by substituting a Sr$^{+2}$ ion for one of the La$^{+3}$ ions in the La$_{2-x}$Sr$_{x}$CuO$_{4}$ (LSCO) compound studied in this work. However, changing $x$ by chemical substitution implies that the microstructure unavoidably varies from sample to sample. Because of the $d$-wave symmetry of the order parameter in copper-oxide superconductors, disorder effects may then become relevant and,  in thin films, even very pronounced, thereby making the interpretation of experiments probing the doping dependence of various quantities a difficult task.  An elegant way to overcome this difficulty relies on the electric-field effect \cite{ATM}, which allows to modulate the free-carrier concentration in the superconducting electrode of a capacitor-like heterostructure by varying the electrostatic field applied across a gate insulator. So far, tunable electric-field-effect devices incorporating a copper-oxide superconductor were almost exclusively studied with conventional transport measurements probing the resitive region above the critical temperature $T_{c}$ \cite{MBMS, XLDKDFV, MBCAHM, MGT}.\\
\indent A powerful method to investigate the superconducting state
of thin films from the fluctuation-dominated region near $T_{c}$
down to \mbox{$T\ll T_{c}$} is to measure their sheet kinetic
inductance $L_{k}$ \cite{JGRLM}. In a two-dimensional (2D)
superconductor $L_{k}^{-1}(T)$ is proportional to the
temperature-dependent areal  superfluid density
 $n_{s\Box}(T)$: \mbox{$L_{k}^{-1}(T)=e^{2}n_{s\Box}(T)/m^{*}$}, where $m^{*}$ is the carrier effective mass. Although not all the free carriers may participate to the superconducting condensate, one can nevertheless expect that varying their areal concentration $n_{\Box}$ will somehow affect $n_{s\Box}$ and, consequently, $L_{k}^{-1}$. Changes of $L_{k}^{-1}$ reflecting changes of $n_{s\Box}$ were indeed observed by Fiory \textit{et al.} \cite{FHEMHO} in the first electric-field-effect experiment performed on a copper-oxide film. However, since the film thickness was much larger than the Thomas-Fermi charge screening length, the relative changes $\Delta L_{k}^{-1}/L_{k}^{-1}$ turned out to be extremely small, typically of the order of $\sim10^{-5}$. Field-induced relative changes of $L_{k}^{-1}$ in the range \mbox{10$^{-4}-10^{-3}$} were also observed in surface impedance measurements at microwave frequencies \cite{FDALXV}, whose interpretation, however, was less transparent because of the non-linear response of the SrTiO$_{3}$ gate insulator to the applied electric field.  \\
\indent In this Letter we describe an experiment in which the
temperature and electric-field dependences of  $L_{k}^{-1}$ were
investigated by capacitively charging an epitaxially grown
ultrathin (two-unit-cell-thick) LSCO film in the underdoped regime
($x=0.1$) with an  electrostatic field $E$ applied across a gate
insulator with a high dielectric constant.
We find that the field-induced change $\Delta L_{k}^{-1}(T, E)$ is a non-monotonic function of $T$ which, for the highest fields accessible to the experiment (\mbox{$E\approx 2\times10^{8}$ V/m}), reaches a maximum corresponding to $\sim20\%$ of  $L_{k}^{-1}(0, 0)$ at $T/T_{c}\approx$ 0.7 and saturates to $\sim10\%$ of $L_{k}^{-1}(0,0)$ at very low temperatures. These large modulations offer new interesting opportunities to explore the intriguing superconducting behavior of the copper oxides. As an illustration, the field-induced relative changes $\Delta L_{k}^{-1}(0, E)/L_{k}^{-1}(0, 0)$ measured at very low temperature were correlated, for the first time on the same sample,  with the corresponding relative variations $\Delta T_{c}(E)/T_{c}(0)$. Both quantities were found to vary linearly with $E$ and, quite remarkably, $\Delta L_{k}^{-1}(0, E)/L_{k}^{-1}(0, 0)=\Delta T_{c}(E)/T_{c}(0)$, a result providing direct compelling evidence for the validity of Uemura's relation \cite{U} $T_{c}\propto n_{s\Box}(0)/m^{*}$ for underdoped copper-oxide superconductors. We consider this observation as the central result emerging from this work. \\
\indent To achieve a substantial electrostatic modulation of
$n_{s\Box}$ in a thin film, it is essential that its thickness $d$
is of the order of the Thomas-Fermi charge screening length
$\lambda_{TF}$, the field-induced charges being confined in a
layer of thickness $\sim\lambda_{TF}$ near the
superconductor-insulator interface. If one relies on a 2D
free-electron gas to estimate $\lambda_{TF}$ for metallic LSCO in
the quasi-2D underdoped regime, one finds
$\lambda_{TF}=\phi_{0}(\epsilon_{s}\epsilon_{0}d_{s}/\pi
m_{e})^{1/2}$, where $d_{s}$ is the CuO$_{2}$-interlayer distance
and $\epsilon_{s}$ the dielectric constant of LSCO ($\phi_{0}$ is
the superconducting magnetic flux quantum). Using $d_{s}=0.66$ nm
and $\epsilon_{s}\approx29$ \cite{CBKPT} for LSCO, one obtains
$\lambda_{TF}\approx0.5$ nm. This means preparing epitaxial films
only a few unit-cells thick in the $c$-axis direction, one
unit-cell (UC) corresponding to a thickness of 1.33 nm. Even when
grown on an epitaxially optimized substrate like SrLaAlO$_{4}$
(SLAO) \cite{LPFMST}, such ultrathin LSCO films have a strongly
reduced $T_{c}$, superconductivity being completely suppressed in
1-UC-thick films. We have recently demonstrated \cite{RCGLFLM},
however, that it is possible to grow with block-by-block molecular
beam epitaxy \cite{JPL} (1-2)-UC-thick LSCO films having $T_{c}$
in the range \mbox{(10-20) K} by inserting a \textit{normal}
(\textit{i.e.}, non-superconducting) metallic LSCO buffer layer
between the superconducting film and the SLAO substrate, a
procedure which apparently helps minimizing the degradation of the
film structure at the interface. Being confined to a region of
thickness $\xi_{c}$ (the $c$-axis coherence length) near the
superconductor-normal metal contact, proximity effects weakening
superconductivity in the LSCO film are expected to be irrelevant
for the 2-UC-thick (\mbox{$d=2.66$} nm) LSCO film studied in this
work, since $\xi_{c}\ll d$ (typically, $\xi_{c}\approx0.1-0.2$ nm
for LSCO in the quasi-2D underdoped regime). Relying on this
almost homoepitaxial method, in a first step a trilayer
heterostructure consisting of a 12-UC-thick normal LSCO buffer
layer ($x=0.4$), a 2-UC-thick superconducting LSCO film in the
underdoped regime ($x=0.1$), and an amorphous  HfO$_{2}$-film of
thickness $D\approx15$ nm \cite {RL} was grown \textit{in situ} on
a monocrystalline SLAO substrate.
\begin{figure}[b]
\begin{center}
\hspace*{-0cm}\vspace{-0.5cm}
\includegraphics[width=8.6cm]{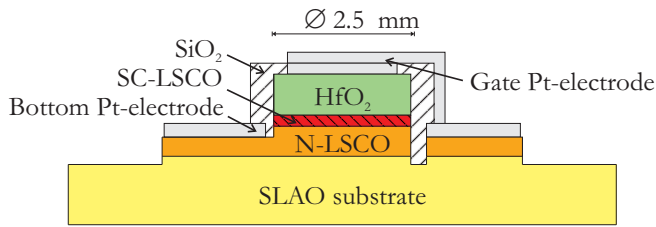} \caption{\label{fig1} Cross section of the multilayer electric-field-effect device. N: Normal, SC: Superconducting.}
\end{center}
\end{figure}
Hafnium oxide was chosen as gate
insulator on account of its high static dielectric constant
$\epsilon$, for which values up to $\epsilon\approx25$ are
reported \cite{R}, and its large breakdown fields, typically
$(6-8)\times10^{8}$ V/m \cite{BMMBG}. In five subsequent
photolithographic steps the trilayer was then patterned in the
capacitor-like "mesa structure" shown in \mbox{Fig. 1}. To avoid
short-circuits, before depositing the top platinum (Pt) gate
electrode, a SiO$_{2}$ layer was deposited around the mesa after
the Pt-metallization of the bottom electrodes. In the gate-voltage
range \mbox{$|V_{G}|\leq$ 3 V} covered by the experiments,
corresponding to electrostatic fields \mbox{$|E|=(|V_{G}|/D)\leq
(2\times10^{8}$) V/m}, the leakage current was found to be
independent of polarity, the maximum leakage current density being
$\sim10^{-4}$ A/cm$^{2}$ at $V_{G}=\pm3$ V. The inverse sheet kinetic inductance $L_{k}^{-1}$ of the superconducting LSCO film was extracted from measurements (down to $T\approx0.5$ K) of the mutual-inductance change of a drive-receive two-coil system caused by the screening  supercurrents flowing in the film in response to a small ac excitation at a frequency of 33 kHz \cite{JGRLM}. \\
\begin{figure}[t]
\begin{center}
\hspace*{-0cm}
\includegraphics[width=8.6cm]{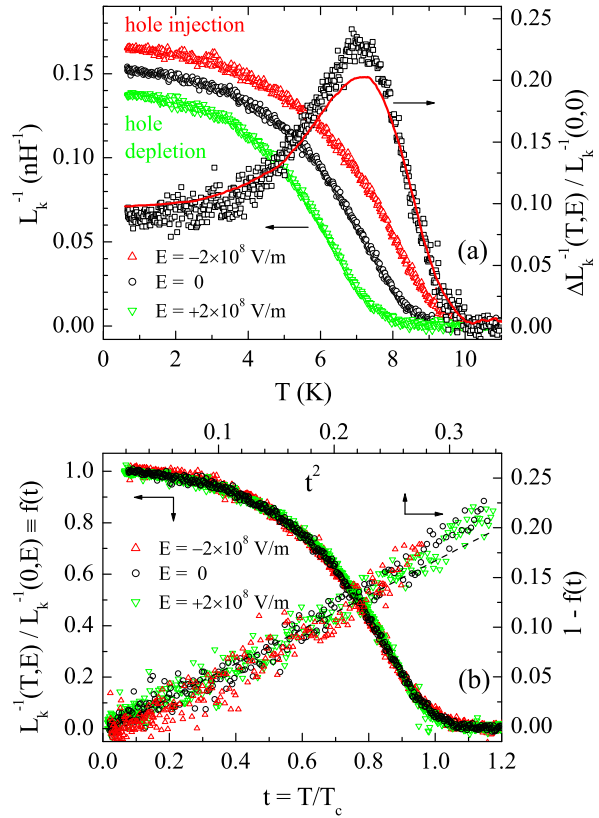}
\vspace{-1cm}\caption{\label{fig2} (a) Temperature dependence of
the inverse kinetic inductance at three values of the gate voltage
\mbox{($V_{G}=0,\pm3$ V)} and of its normalized deviation for
$|V_{G}|=3$ V from  the $V_{G}=0$ data of an underdoped ($x=0.1$)
2-UC-thick LSCO film. The full curve is a fit according to Eq.
(3); (b) Universal temperature dependence $f(t)$ of the normalized
$L_k^{-1}$ vs $T$ curves of (a). Notice the $t^2$-dependence of
$[1-f(t)]$ at low temperatures ($t<$0.45) highlighted by the
dashed straight line.}
\end{center}
\end{figure}
\indent In \mbox {Fig. 2(a)} the temperature dependence of
$L_{k}^{-1}$ is shown for \mbox{$V_{G}=0, \pm3$} V. The data
reveal a marked electrostatic modulation of $n_{s\Box}(T)$ over
the whole temperature range, $L_{k}^{-1}(T)$ exhibiting a
substantial enhancement for hole injection (corresponding to
$V_{G}<0$) and an almost identical reduction for hole depletion
(corresponding to $V_{G}>0$). This is qualitatively consistent
with the fact that the LSCO film is in the underdoped regime,
where $n_{s\Box}(T)$ is expected to rise with increasing hole
concentration. When plotted against the reduced temperature
$t\equiv T/T_{c}(E)$, where $T_{c}(E)$ was defined \textit{ad hoc}
by extrapolating to zero the linear high-temperature portion of
the  $L_{k}^{-1}$ vs $T$ curves, the normalized data
$L_{k}^{-1}(T, E)/L_{k}^{-1}(0, E)$ all collapse on the universal
curve $f(t)$ shown in Fig. 2(b). This demonstrates that, at least
in the limited doping range explored in this work,
$f(t)$ is independent of the carrier concentration, a conclusion unaffected by the choice of the procedure adopted to define $T_{c}(E)$. In other words, within the framework of a two-fluid model of superconductivity, this means that the ratio $n_{n\Box}(T)/n_{s\Box}(T)$ between the areal normal-fluid density $n_{n\Box}(T)$ of the thermally excited quasiparticles and $n_{s\Box}(T)$ is independent of doping. In Fig. 2(b) we also show that, at low temperatures, $[1-f(t)]$, which is proportional to the change of $n_{n\Box}(T)$ with temperature, follows the $t^{2}$-dependence predicted by the theory of disordered $d$-wave superconductors \cite{HG}. \\
\indent Deeper insight into the nature of the superconducting
state of the copper oxides in the underdoped regime is provided by
studying the relationship between the field-induced relative shift
$\Delta T_{c}(E)/T_{c}(0)$ of the transition temperature and the
relative change $\Delta L_{k}^{-1}(0, E)/L_{k}^{-1}(0, 0)$ of the
inverse kinetic inductance at very low temperature, in the limit
$T\rightarrow0$. As shown in \mbox{Fig. 3}, both quantities vary
linearly with $E$ in the field-range $E=\pm2\times10^{8}$ V/m.
Most remarkable, however, is the observation that, within
experimental accuracy,  $\Delta L_{k}^{-1}(0, E)/L_{k}^{-1}(0,
0)=\Delta T_{c}(E)/T_{c}(0)$, a result pointing unambiguously, at
least in the limited carrier concentration range covered by our
experiment, to the relation $T_{c}\propto n_{s\Box}(0)/m^{*}$
proposed by Uemura \textit{et al.} \cite{U}, but in striking
contrast with recent $L_{k}$-measurements performed on underdoped
YBCO films, where $T_{c}$ was found roughly proportional to
$[n_{s\Box}(0)/m^{*}]^{1/2}$ \cite{ZKL}. We think that this
discrepancy might result from sample-to-sample varying disorder in
the experiments of Ref. \cite{ZKL}, although other explanations
can be envisaged. Agreement of our result with a recently proposed
alternative scaling relation \cite{H} would imply that the
normal-state conductivity at $T_{c}$ is independent of $n_{\Box}$,
which is definitely not the case \cite{HS}.
We emphasize that the message conveyed by our observation is clearly distinct from that emerging from the field-effect experiments of Ref. \cite{XDWKLV}, in which the relation $T_{c}\propto n_{\Box}$ was inferred from transport measurements. As a matter of fact, $n_{s\Box}(0)$ and $n_{\Box}$ are likely to be related, but not necessarily identical, quantities. \\
\indent At this point, it is instructive to compare our results
with existing theoretical expressions relating $L_{k}^{-1}(0)$ to
$T_{c}$. Assuming $n_{s\Box}(0)=n_{\Box}$, the quasiparticle approach \cite{LW} predicts $L_{k}^{-1}(0)/T_{c}=(k_{B}e^{2}/m^{*}a^{2})(3J/t_{h})\Delta_{0}^{-1}$. Typical values for cuprates are $J\approx130$ meV for the exchange energy and $J/t_{h}\approx0.3$ for the ratio between $J$ and the tight-binding hopping integral $t_{h}$ \cite{LW}. Then, since the maximum value $\Delta_{0}$ of the $d$-wave energy gap is roughly $J/3\approx 45$ meV \cite{LW}, $L_{k}^{-1}(0)/T_{c}\approx 0.2$ nH$^{-1}$/K, using $m^{*}\approx 2m_{e}$ \cite{LW} and \mbox{$a=0.38$ nm} for the cell side of the square
CuO$_{2}$ lattice. An alternative prediction, based on a phase-fluctuation mechanism \cite{EK, S}, leads, in two dimensions, to $L_{k}^{-1}(0)/T_{c}=Ank_{B}(2\pi/\phi_{0})^{2}$, where $A$ is a constant of order unity and $n$, the number of CuO$_{2}$ planes in the LSCO film, reflects the fact that the measured $L_{k}$ is the parallel connection of the kinetic inductances of the individual CuO$_{2}$ planes. For our 2-UC-thick LSCO film ($n=4$) this gives $L_{k}^{-1}(0)/T_{c}\approx 0.5$ nH$^{-1}$/K. Both theoretical scenarios lead to values which are at least one order of magnitude larger than that actually observed in our experiment [$L_{k}^{-1}(0)/T_{c}\approx 0.02$ nH$^{-1}$/K]. This strongly suggests that disorder not
only considerably reduces $n_{s\Box}(0)$, but also tends to
suppress $n_{s\Box}(0)$ more efficiently than $T_{c}$, however
without altering the proportionality between $L_{k}^{-1}(0)$ and
$T_{c}$. In this connection, we notice that the different
sensitivity of  $L_{k}^{-1}(0)$ and $T_{c}$ to disorder is consistent with previous experimental observations and can be explained by a mean-field theory of $d$-wave superconductors in which the order-parameter spatial variations caused by disorder in short-coherence-length materials are adequately taken into account \cite{FKBS}. For the very same reason (\textit{i.e.}, $T_{c}$ more robust against disorder than $n_{s\Box}$), we argue that the universal "superfluid jump" at $T_{c}$ predicted by the Berezinskii-Kosterlitz-Thouless theory tends to be suppressed by disorder, a conjecture consistent with the absence of any rapid superfluid drop near $T_{c}$ in the $L_{k}^{-1}$ vs $T$ curves of Fig. 2. \\
\begin{figure}[tb]
\begin{center}
\hspace*{-0cm}
\includegraphics[width=8.6cm]{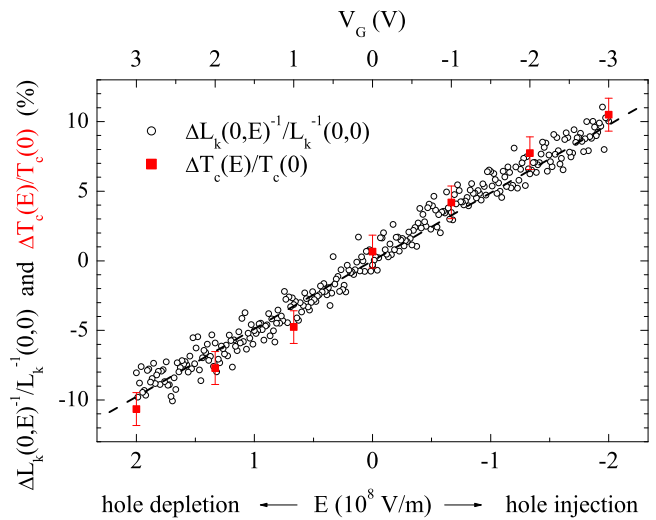}
\vspace{-0.5cm} \caption{\label{fig3} Comparison of the
field-induced relative changes of the critical temperature and of
the inverse kinetic inductance at very low temperature ($T=0.6$ K,
$T/T_{c}\approx 0.07$) for an underdoped ($x=0.1$) 2-UC-thick LSCO
film as a function of the applied electrostatic field. The dashed
straight line is a fit according to Eq. (2).}
\end{center}
\end{figure}
\indent In order to explain the $\Delta L_{k}^{-1}(0,
E)/L_{k}^{-1}(0, 0)$ data of \mbox{Fig. 3} in more quantitative
terms, we make the reasonable assumption that the fraction
\mbox{$n_{s\Box}(0)/n_{\Box}$} of holes participating to the
superconducting condensate at \mbox{$T=0$} is independent of
their concentration and, consequently, of $E$. Then, $\Delta
L_{k}^{-1}(0, E)/L_{k}^{-1}(0, 0)=\Delta n_{\Box}(E)/n_{\Box}$. To
determine the field-induced change $\Delta n_{\Box}(E)$ of the
areal hole density $n_{\Box}$, we ignore the discrete layered
structure of LSCO and assume that it behaves like a 2D metallic
continuum, whose charge screening properties are described by the
2D expression of $\lambda_{TF}$ given above. Relying on this
simple model, elementary electrostatics leads to:
\begin{equation}\label{1} \Delta
n_{\Box}(E)=(CD/e)[1-\exp{(-d/\lambda_{TF})}]E,
\end{equation} where \mbox{$C^{-1}=(\epsilon\epsilon_{0}/D)^{-1}+(\epsilon_{0}/\lambda_{TF})^{-1}$}, an expression showing that the capacitance $C$ (per unit surface) of the field-effect structure is the series connection of the "geometrical" capacitance $\epsilon\epsilon_{0}/D$ and the "interface" capacitance $\epsilon_{0}/\lambda_{TF}$. It should be noticed, however, that other contributions to the interface capacitance (see, for instance, Ref. \cite{SBDH})
are not included in this treatment. In our experiment,
$d/\lambda_{TF}\approx 5$ and it turns out, \textit{a posteriori},
that $\lambda_{TF}\lesssim D/\epsilon$, so that:
\begin{equation}\label{2} \Delta L_{k}^{-1}(0, E)/L_{k}^{-1}(0,
0)=\epsilon\epsilon_{0}E/en_{\Box}.
\end{equation}
Expressing $n_{\Box}$ in terms of the Sr-concentration
[$n_{\Box}=(x/a^{2}d_{s})d=2.78\times10^{18}$ m$^{-2}$], Eq. (2) can be fitted to the data of
\mbox{Fig. 3} taking $\epsilon\approx 24$, a value in good
agreement with that reported for HfO$_{2}$ in Ref. \cite{R}.
Notice, incidentally, that in deriving Eq. (1) the dielectric constant of HfO$_{2}$ was
assumed to be independent of $E$, which is consistent with the linear field dependence of the
data of Fig. 3.\\
\indent Finally, Uemura's relation $L_{k}^{-1}(0, E)\propto
T_{c}(E)$ and the universality of $f(t)$, the two basic results of
this work, can be combined to describe the temperature dependence
of the field-induced change $\Delta L_{k}^{-1}(T, E)$. Expanding
\mbox{$L_{k}^{-1}(T, E)=L_{k}^{-1}(0, E)f[T/T(E)]$} about
\mbox{$E=0$}, one obtains:
\begin{equation}\label{3}
\frac{\Delta L_{k}^{-1}(T, E)}{L_{k}^{-1}(0, 0)}=\left [f(t)-t\frac{df}{dt}\right]\frac{\Delta T_{c}(E)}{T_{c}(0)},
\end{equation}
where \mbox{$t=T/T_{c}(0)$}. In Fig. 2(a) the temperature dependence of $\Delta L_{k}^{-1}(T, E)/L_{k}^{-1}(0, 0)$, as deduced by averaging the deviations of $L_{k}^{-1}(T,E)$ (corresponding to $E=\pm(2\times10^{8}$) V/m) with respect to $L_{k}^{-1}(T,0)$, is compared to Eq. (3), where the expression in the brackets was calculated from the universal function $f(t)$ shown in Fig. 2(b). The agreement is very good, thereby demonstrating the overall consistency of our observations.\\
\indent In conclusion, large electrostatic modulations of the
areal superfluid density $n_{s\Box}$ of an ultrathin LSCO film in
the underdoped regime were observed by charging the film in an
accurately devised capacitor structure. The central result
emerging from our investigations is that  the proportionality
between $T_{c}$ and $n_{s\Box}(T=0)$, empirically proposed by
Uemura, was verified, for the first time, on the same sample and
is thus unaffected by the uncertainties resulting from
sample-to-sample varying disorder of other experimental
approaches. When adequately normalized, the $n_{s\Box}(T,E)$ data
exhibit a universal temperature dependence, which, at low
temperatures, follows the theoretical prediction for disordered
$d$-wave superconductors.\\

\indent We
thank H. Beck, A.M. Goldman, A.F. Hebard, S.E. Korshunov, A.E.
Koshelev,  J. Mannhart, and J.-M. Triscone for clarifying
discussions and H. Siegwart for technical assistance. This work
was supported by the Swiss National Science Foundation through the
National Center of Competence in Research "Materials with Novel
Electronic Properties".

\end{document}